\documentclass{iaus}
\usepackage{graphicx}

\title[Lithium in the Globular Cluster NGC 6397] 
{Lithium abundances of Main Sequence and Sub-Giant stars in the Globular Cluster NGC 6397}

\author[Gonz\'alez Hern\'andez et al.]   
{J. I. Gonz\'alez Hern\'andez$^{1,2}$\thanks{Present address: Dpto. de Astrof\'{\i}sica y Ciencias de la Atm\'osfera, Facultad de
F\'{\i}sica, Universidad Complutense de Madrid, E-28040 Madrid,
Spain. Email: {\tt jonay@astrax.fis.ucm.es}}, P. Bonifacio$^{1,2,3}$,
E. Caffau$^1$, M. Steffen$^4$, \\  H.-G. Ludwig$^{1,2}$, N.
Behara$^{1,2}$, L. Sbordone$^{1,2}$, R. Cayrel$^1$, \and S.~Zaggia$^5$
}

\affiliation{
$^1$GEPI, Observatoire de Paris, CNRS, Universit\'e Paris Diderot; \\[\affilskip]Place
Jules Janssen 92190 Meudon, France \\ email: {\tt
Jonay.Gonzalez-Hernandez@obspm.fr}\\[\affilskip] 
$^2$Cosmological Impact of the First STars (CIFIST) Marie Curie Excellence
Team\\[\affilskip]
$^3$Istituto Nazionale di Astrofisica - Observatorio\\[\affilskip] 
Astronomico di Trieste, Italy\\[\affilskip]
$^4$Astrophysikalisches Institut Potsdam, An der Sternwarte 16, \\[\affilskip] 
D-14482 Potsdam, Germany \\[\affilskip]
$^5$INAF - Osservatorio Astronomico di Padova, \\[\affilskip] 
Vicolo dell'Osservatorio 5, Padua 35122, Italy\\[\affilskip] 
}

\pubyear{2009}
\volume{266}  
\pagerange{xxx--xxx}
\date{30 September 2009 and in revised form 30 September 2009}
\jname{Star Clusters: Basic Galactic Building Blocks Throughout Time And Space}
\editors{Richard de Grijs \& Jacques Lepine, eds.}
\begin{document}

\maketitle

\begin{abstract}

We present FLAMES/GIRAFFE spectroscopy obtained at the Very Large
Telescope (VLT). Using these observations we have been able for the
first time to observe the Li I doublet in the Main Sequence stars of a
Globular Cluster. We also observed Li in a sample of
Sub-Giant stars of the same B-V colour. 

Our final sample is composed of 84 SG stars and 79 MS stars.
In spite of the fact that SG and MS span the same temperature range we
find that the equivalent widths of the Li I doublet in SG stars are 
systematically larger than those in MS stars, suggesting a higher Li
content among SG stars. This is confirmed by our quantitative analysis
which makes use of both 1D and 3D model atmospheres.

We find that SG stars show, on average, a Li abundance higher by
0.1\,dex than MS stars. We also detect a positive slope of Li 
abundance with effective temperature, the higher the temperature the 
higher the Li abundance, both for SG and MS stars, although the
slope is slightly steeper for MS stars. These results provide an
unambiguous evidence that the Li abundance changes with evolutionary
status.  

The physical mechanisms that contribute to this are not yet clear,
since none of the proposed models seems to describe accurately the
observations. Whether such mechanism can explain the cosmological
lithium problem, is still an open question.

\end{abstract}

\firstsection 
\section{Introduction}

The primordial Li abundance inferred from the fluctuations of cosmic
microwave background measured by the WMAP satellite
(\cite[Spergel et al. 2007, Cyburt et al. 2008]{spe07,cyburt}) is $\log ({\rm Li}/{\rm H})+ 12 =2.72\pm0.06$,
approximately 0.3--0.5 dex higher than the Li abundance determined in
metal-poor stars of the Galactic halo.

Many studies have tried to explain this difference: 
(a) \cite[Piau et al. 2006]{pia06} propose that the first generation
of stars, Population III stars, could have processed some fraction of
the halo gas, lowering the lithium abundance;
(b) other authors suggest that the primordial Li abundance has been
uniformly depleted in the atmospheres of metal-poor dwarfs by some
physical mechanism (e.g. turbulent diffusion as in \cite{ric05,kor06};
gravitational waves as in \cite{cat05}, etc.); and (c) finally, it has
been also suggested that the standard Big Bang nucleosynthesis (SBBN)
calculations should be revised, possibly with the introduction of new
physics as in e.g. \cite{jed04,jed06,jittoh,hisano}. 

Here we present the Li abundances of subgiant (SG) and main-sequence
(MS) stars of the cluster NGC~6397. This study represents the first Li
abundance measurements in MS stars of a Globular Cluster. 

\section{Observations}

Spectroscopic observations of the globular cluster NGC~6397 
were carried out with the multi-object spectrograph FLAMES-GIRAFFE at
the VLT on 2007 April, May, June and July, covering the spectral
range $\lambda\lambda$6400--6800 {\AA} at resolving power
$\lambda/\delta\lambda\sim17,000$.  

We selected subgiant and dwarf stars in the colour range
$B-V=0.6\pm0.03$, which ensures that both set of stars fall in a
similar and narrow effective temperature range (see Fig.~3 online in
\cite[Gonz\'alez Hern\'andez et al. 2009]{gon09}).

The spectra were reduced with the ESO pipeline and later on treated
within MIDAS. We correct the spectra for sky lines and, barycentric
and radial velocity. We typically combine 17 spectra of dwarfs and 4
spectra of subgiants to achieve a similar S/N ratio. The mean radial
velocity of the cluster stars is $v_r=18.5$~km~s$^{-1}$.

\section{Stellar parameters}

We derived the effective temperature by fitting the observed
H$\alpha$ line profiles with synthetic profiles, using 3D
hydrodynamical model atmospheres computed with the CO$^5$BOLD code.
The details of this code are provided in \cite{fre02,wed04}. 
The ability of 3D models to reproduce Balmer line profiles has been
shown in \cite{beh09}, where the H$\alpha$ profiles of the Sun, and
the metal-poor stars HD 84937, HD 74000 and HD 140283 were
investigated. See also \cite{lud09} for further details. 

The effective temperatures of MS and SG stars were also derived
adopting 1D ATLAS~9 model atmospheres (see \cite[Kurucz 2005]{kur05})
and using the same fitting procedure. 

Fixed values for the surface gravity were adopted for both subgiant
and dwarf stars in the sample, according to the values that best match
the position of the stars on a 12 Gyr isochrone (\cite[Straniero et
al. 1997]{str97}). The adopted values were $\log (g/{\rm
cm~s}^2)=4.40$ and 3.85 for MS and SG stars, respectively. 

\section{Li abundances}

We measure the equivalent width (EW) of the Li~{\scriptsize
I}~6708~{\AA} line in SG and MS stars by fitting synthetic spectra of
known EW to the observed Li profiles. 
The SG stars show on average larger EWs than MS stars. This is clearly
seen in Fig.~1 of \cite{gon09}, where the histograms of the EWs
measured in SG and MS stars of this cluster are displayed. Although
the colour $B-V$ is sensitive to surface gravity, a priori, this
result was not expected, and suggest that subgiants in this cluster
have actually higher Li abundances than dwarfs. 

\begin{figure}
\centering
\includegraphics[width=8.3cm,angle=0]{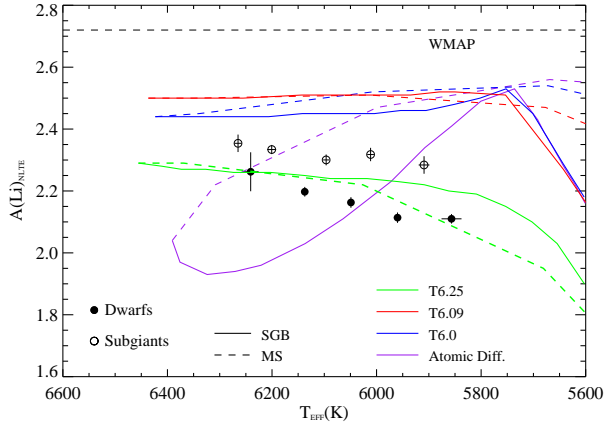}
\caption{\footnotesize{1D NLTE Li abundances versus 1D effective
temperatures of the observed stars together with Li isochrones for
different turbulent diffusion models. Black filled circles and open circles
correspond to dwarf and subgiant stars, respectively.  
Different turbulent diffusion model isochrones of lithium are
displayed for different levels of turbulence. In each isochrone, the
dashed stretch of the line shows the Li abundance in dwarf stars,
whereas the solid line shows the Li abundance in subgiant stars. The
horizontal black dashed line depicts the cosmological Li abundance.}}
\label{figali} 
\end{figure}

The Li abundances were derived using 3D model atmospheres. The line
formation of Li was treated in non-local thermodynamical equilibrium
(NLTE) using the same code and model atom used in \cite{cay07}. 
The analysis has been also done with 1D model
atmospheres providing essentially the same picture, although
$T_{\rm eff}$ in 1D show lower values (see Fig.~6 online in
\cite[Gonz\'alez Hern\'andez et al. 2009]{gon09}).
In the 1D case we use the \cite{carlsson} NLTE corrections.

\section{Discussion and Conclusions}

In Fig.~\ref{figali} we display the derived 1D-NLTE Li 
abundances versus the 1D effective temperatures of dwarf and 
subgiant stars of the globular cluster NGC 6397 (see Fig.~2 in
\cite{gon09} for a similar picture
but with $T_{\rm eff}$ and Li abundances computed using 3D models).
The stars have been divided into five bins of effective temperature.
The error bar in the Li abundance is the dispersion for each bin,
divided by the square root of the number of stars in the bin.
The Li abundance decreases with decreasing temperature.
This lithium abundance pattern is different from what is
found among field stars (\cite[Mel\'endez \& Ram{\'\i}rez 2004]{mel04},
\cite[Bonifacio et al. 2007]{bon07}, \cite[Gonz\'alez Hern\'andez et al. 2008]{gon08}). 

The difference in the Li abundance of dwarfs and subgiants is
$\sim0.14$~dex if one computes the mean 1D A(Li) and the standard
deviation of the mean for the two samples. For subgiants we find mean
1D Li abundances of $2.31\pm0.01$, while for dwarfs $2.17\pm0.01$. The
3D models provide similar results although slightly higher Li
abundances. 
\cite{lin09} also find such difference 
between the mean Li abundances in MSs and SGs, but only by 0.03~dex
although still significant at 1$\sigma$. However, this result is partly
affected by the very narrow range of $T_{\rm eff}$ for MSs deduced by
\cite{lin09} ($\sim 80$~K) compared to the wide range 
($\sim 450$~K) for the SGs (see Fig.~7 online in
\cite[Gonz\'alez Hern\'andez et al. 2009]{gon09}).

Our results imply that the Li surface abundance depends on the
evolutionary status of the star.
In Fig.~\ref{figali} we show the Li isochrones for
different turbulent diffusion models (\cite[Richard et al.
2005]{ric05}). These models have
been shifted up by 0.14 dex in Li abundance to make the initial
abundance of the models, $\log ({\rm Li}/{\rm H})=2.58$, coincide
with the primordial Li abundance predicted from fluctuations of the
microwave background measured by the WMAP satellite (\cite[Cyburt et
al. 2008]{cyburt}). 
The models assuming pure atomic diffusion, and, among those including
turbulent mixing, T6.0 and T6.09, are ruled out by our observations. 
All such models predict that in dwarf stars Li
should be either more abundant or the same as in subgiant stars.
The only model which predicts a Li pattern which is qualitatively
similar to that observed, is the T6.25 model.

Models including atomic diffusion and tachocline 
mixing (\cite[Piau 2008]{pia08}) do not seem to reproduce our observations, since
they provide a constant Li abundance up to 5500\,K. The sophisticated
models which besides diffusion and rotation also take into account the
effect of internal gravity waves (\cite[Talon \& Charbonnel
2004]{tac04}), seem to predict accurately the Li abundance pattern in
solar-type stars, at solar metallicity (\cite[Charbonnel \& Talon
2005]{cat05}), but models at low metallicity are still needed. 

The cosmological lithium discrepancy still needs to be solved.
Given that none of the existing models of Li evolution
in stellar atmospheres matches the observations, we hope
our results will prompt further new theoretical investigations.


\begin{thebibliography}{}

\bibitem[Behara et al.(2009)]{beh09} 
Behara, N.~T., Ludwig, H.-G., Steffen, M., \& Bonifacio, P.\ 2009,
\textit{American Institute of Physics Conference Series}, 1094, 784 

\bibitem[Bonifacio et al.(2007)]{bon07} 
Bonifacio, P., et al.\ 2007, \textit{A\&A}, 462, 851 

\bibitem[Carlsson et al.(1994)]{carlsson} Carlsson, M., 
Rutten, R.~J., Bruls, J.~H.~M.~J., \& Shchukina, N.~G.\ 1994,
\textit{A\&A}, 288, 860  

\bibitem[Cayrel et al.(2007)]{cay07} 
Cayrel, R., et al.\ 2007, \textit{A\&A}, 473, L37 

\bibitem[Charbonnel \& Talon(2005)]{cat05} 
Charbonnel, C., \& Talon, S.\ 2005, \textit{Science}, 309, 2189 

\bibitem[Cyburt et al.(2008)]{cyburt}
Cyburt, R.~H., Fields, B.~D., \& Olive, K.~A.\ 2008, \textit{Journal of
Cosmology and Astro-Particle Physics}, 11, 12  

\bibitem[Freytag et al.(2002)]{fre02} 
Freytag, B., Steffen, M., \& Dorch, B.\ 2002, \textit{Astronomische
Nachrichten}, 323, 213  

\bibitem[Gonz{\'a}lez Hern{\'a}ndez et al.(2008)]{gon08} 
Gonz{\'a}lez Hern{\'a}ndez, J.~I., et al.\ 2008, \textit{A\&A}, 480, 233 

\bibitem[Gonz{\'a}lez Hern{\'a}ndez et al.(2009)]{gon09} 
Gonz{\'a}lez Hern{\'a}ndez, J.~I., et al.\ 2009, \textit{A\&A Letters
}, 505, L13

\bibitem[Hisano et al.(2009)]{hisano} 
Hisano, J., Kawasaki, M., Kohri, K., \& Nakayama, K.\ 2009,
\textit{Phys. Rev. D.}, 79, 063514  

\bibitem[Jedamzik(2004)]{jed04} 
Jedamzik, K.\ 2004, \textit{Phys. Rev. D.}, 70, 083510 

\bibitem[Jedamzik(2006)]{jed06} 
Jedamzik, K.\ 2006, \textit{Phys. Rev. D.}, 74, 103509 

\bibitem[Jittoh et al.(2008)]{jittoh} 
Jittoh, T., Kohri, K., Koike, M., Sato, J., Shimomura, T., \&
Yamanaka, M.\ 2008, \textit{Phys. Rev. D.}, 78, 055007  

\bibitem[{{Kurucz}(2005)}]{kur05}
{Kurucz}, R.~L. 2005, \textit{Memorie della Societ\`a
Astronomica Italiana Supplement}, 8, 14

\bibitem[Korn et al.(2006)]{kor06} 
Korn, A.~J., Grundahl, F., Richard, O., Barklem, P.~S., Mashonkina,
L., Collet, R., Piskunov, N., \& Gustafsson, B.\ 2006, \textit{Nature}, 442, 657

\bibitem[Lind et al.(2009)]{lin09} 
Lind, K., Primas, F., Charbonnel, C., Grundahl, F., \& Asplund, M.\
2009, \textit{A\&A}, 503, 545  

\bibitem[Ludwig et al.(2009)]{lud09} 
Ludwig, H.-G., Behara, N.~T., Steffen, M., \& Bonifacio, P.\ 2009,
\textit{A\&A}, 502, L1  

\bibitem[Mel{\'e}ndez \& Ram{\'{\i}}rez(2004)]{mel04} 
Mel{\'e}ndez, J., \& Ram{\'{\i}}rez, I.\ 2004, \textit{ApJ}, 615, L33 

\bibitem[Piau et al.(2006)]{pia06} 
Piau, L., Beers, T.~C., Balsara, D.~S., Sivarani, T., Truran, J.~W., 
\& Ferguson, J.~W.\ 2006, \textit{ApJ}, 653, 300 

\bibitem[Piau(2008)]{pia08} 
Piau, L.\ 2008, \textit{ApJ}, 689, 1279 

\bibitem[Richard et al.(2005)]{ric05} 
Richard, O., Michaud, G., \& Richer, J.\ 2005, \textit{ApJ}, 619, 538 

\bibitem[Spergel et al.(2007)]{spe07} 
Spergel, D.~N., et al.\ 2007, \textit{ApJS}, 170, 377 

\bibitem[Straniero et al.(1997)]{str97} 
Straniero, O., Chieffi, A., \& Limongi, M.\ 1997, \textit{ApJ}, 490, 425 

\bibitem[Talon \& Charbonnel(2004)]{tac04} 
Talon, S., \& Charbonnel, C.\ 2004, \textit{A\&A}, 418, 1051 

\bibitem[Wedemeyer et al.(2004)]{wed04}
Wedemeyer, S., Freytag, B., Steffen, M., Ludwig, H.-G., \& Holweger,
H.\ 2004, \textit{A\&A}, 414, 1121  

\end{thebibliography}
\end{document}